\documentclass[10pt,conference]{IEEEtran}
\IEEEoverridecommandlockouts
\usepackage{dblfloatfix}
\usepackage{setspace}
\usepackage{adjustbox}
\usepackage{cite}
\usepackage{amsmath,amssymb,amsfonts, pifont}
\usepackage{algorithmic}
\usepackage{graphicx}
\usepackage{textcomp}
\usepackage{xcolor}
\usepackage{listings}
\usepackage{color}
\usepackage{anyfontsize}
\usepackage{hyperref}
\newcommand{\cmark}{\ding{51}}%
\newcommand{\xmark}{\ding{55}}%
\usepackage{fourier}
\UseRawInputEncoding

\usepackage{listings, xcolor}

\definecolor{verylightgray}{rgb}{.97,.97,.97}

\lstdefinelanguage{Solidity}{
	keywords=[1]{anonymous, assembly, assert, balance, break, call, callcode, case, catch, class, constant, continue, constructor, contract, debugger, default, delegatecall, delete, do, else, emit, event, experimental, export, external, false, finally, for, function, gas, if, implements, import, in, indexed, instanceof, interface, internal, is, length, library, log0, log1, log2, log3, log4, memory, modifier, new, payable, pragma, private, protected, public, pure, push, require, return, returns, revert, selfdestruct, send, solidity, storage, struct, suicide, super, switch, then, this, throw, transfer, true, try, typeof, using, value, view, while, with, addmod, ecrecover, keccak256, mulmod, ripemd160, sha256, sha3}, 
	keywordstyle=[1]\color{blue}\bfseries,
	keywords=[2]{address, bool, byte, bytes, bytes1, bytes2, bytes3, bytes4, bytes5, bytes6, bytes7, bytes8, bytes9, bytes10, bytes11, bytes12, bytes13, bytes14, bytes15, bytes16, bytes17, bytes18, bytes19, bytes20, bytes21, bytes22, bytes23, bytes24, bytes25, bytes26, bytes27, bytes28, bytes29, bytes30, bytes31, bytes32, enum, int, int8, int16, int24, int32, int40, int48, int56, int64, int72, int80, int88, int96, int104, int112, int120, int128, int136, int144, int152, int160, int168, int176, int184, int192, int200, int208, int216, int224, int232, int240, int248, int256, mapping, string, uint, uint8, uint16, uint24, uint32, uint40, uint48, uint56, uint64, uint72, uint80, uint88, uint96, uint104, uint112, uint120, uint128, uint136, uint144, uint152, uint160, uint168, uint176, uint184, uint192, uint200, uint208, uint216, uint224, uint232, uint240, uint248, uint256, var, void, ether, finney, szabo, wei, days, hours, minutes, seconds, weeks, years},	
	keywordstyle=[2]\color{teal}\bfseries,
	keywords=[3]{block, blockhash, coinbase, difficulty, gaslimit, number, timestamp, msg, data, gas, sender, sig, value, now, tx, gasprice, origin},	
	keywordstyle=[3]\color{violet}\bfseries,
	identifierstyle=\color{black},
	sensitive=false,
	comment=[l]{//},
	morecomment=[s]{/*}{*/},
	commentstyle=\color{gray}\ttfamily,
	stringstyle=\color{red}\ttfamily,
	morestring=[b]',
	morestring=[b]"
}
\usepackage{balance}

\begin{document}
\author{\IEEEauthorblockN{Noama Fatima Samreen, Manar H. Alalfi}
\IEEEauthorblockA{\textit{Department of Computer Science} \\
\textit{Ryerson University,
Toronto, ON, Canada} \\
{\{noama.samreen,manar.alalfi\}@ryerson.ca}}}
\title{{SmartScan: An approach to detect Denial of Service Vulnerability in Ethereum Smart Contracts}
}
\vspace{-0.7 cm}

%
\maketitle
\vspace{-0.3 cm}
\begin{abstract}
Blockchain technology (BT) Ethereum Smart Contracts allows programmable transactions that involve the transfer of monetary assets among peers on a BT network independent of a central authorizing agency. Ethereum Smart Contracts are programs that are deployed as decentralized applications, having the building blocks of the blockchain consensus protocol. This technology enables consumers to make agreements in a transparent and conflict-free environment. However, the security vulnerabilities within these smart contracts are a potential threat to the applications and their consumers and have shown in the past to cause huge financial losses. In this paper, we propose a framework that combines static and dynamic analysis to detect \textit{Denial of Service (DoS) vulnerability  due to an unexpected revert} in Ethereum Smart Contracts.
Our framework, SmartScan,  statically scans smart contracts under test (SCUTs)  to identify patterns that are potentially vulnerable in these SCUTs and then uses dynamic analysis to precisely confirm their exploitability  of the \textit{DoS-Unexpected Revert} vulnerability, thus achieving increased performance and more precise results. We evaluated SmartScan on a set of 500 smart contracts collected from the Etherscan\cite{Etherscsan}. Our approach shows an improvement in precision and recall when compared to available state of the art techniques.


\end{abstract}
\section{Introduction}
Attributing to the wide range applicability of Blockchain technology (BT), it has been finding popularity in many domains. Bitcoin was the first version of cryptocurrency applied using blockchain technology and has since been used in many other applications such as e-commerce, trade and commerce, production and manufacturing, banking, and gaming. BT uses a peer-to-peer framework which is a more decentralized approach to storing transactions and data registers. As there is no single point of failure or a third-party centralized control of transactions, BT has been standing out from other emerging technologies. It uses a chain of blocks in which each block is locked using cryptography  using the hash of the previous block it is linked to, which creates an immutable database of all transactions stored as a digital ledger, and it cannot be changed without affecting all the blocks linked together in the chain\cite{surveyattacks}. 

Proof of Work (POW) and Proof of Stake (POS) are the systems that are required to be performed by each connected miner in a BT network wherein a complex computational math problem is solved to become eligible to add a block to the BT network. Data security and integrity are vital and to ensure protection from unauthorized access, the aspect of using a cryptography-based locked chain of blocks is introduced. Our research is focused on Ethereum Blockchain and the security vulnerabilities in it. Recent research have shown that many applications have been exposed to attacks because of vulnerabilities found in Solidity-based Ethereum Smart Contracts\cite{vulnerabilities}.
One such vulnerability in Solidity Smart Contract is the \textit{DoS-Unexpected Revert} vulnerability. \textit{DoS - Denial of Service}, as the name suggests, is a vulnerability where a Smart Contract is rendered inoperable because of an improper handling of an incomplete transaction resulting from a failure or a deliberate revert of the transaction.
This paper presents a framework to address the \textit{DoS due to Unexpected Revert} attack and provides a framework to identify this vulnerability. This paper has the following contributions:
\begin{enumerate}
    \item An automated framework that combines static and dynamic analysis to better capture the various patterns of the \textit{DoS due to an Unexpected Revert} attack efficiently and accurately in over 500 SCUTs. [See Table \ref{Patterns} for various DOS-Unexpected Revert Vulnerability patterns identified]
\item A Web application to scan Solidity-based Ethereum Smart Contracts for \textit{DOS-Unexpected Revert} vulnerability.
\end{enumerate}
\section{Background}
\vspace{-0.1 cm}
\subsection{Ethereum}
Ethereum\cite{Ethereum} is a BT platform that implements algorithms expressed in a general purpose programming language allowing developers to build a variety of applications, ranging from simple wallet applications to complex financial systems for the banking industry. These programs are known as Smart Contracts which are written in a Turing-complete byte-code language, called EVM byte-code\cite{Ethereum}. The transactions sent to the Ethereum network by the users can create new contracts, invoke functions of a contract, and/or transfer ether to contracts.
\subsection{Smart Contract}
\vspace{-0.1 cm}
Ethereum uses Smart Contracts\cite{Ethereum}, which are computer programs that directly controls the flow or transfer of digital assets.  Each function invocation in a Smart Contract is executed by all miners in the Ethereum network and they receive execution fees paid by the users for it. Execution fees also protect against denial-of-service attacks, where an attacker tries to slow down the network by requesting time-consuming computations. This execution fee is defined as a product of \textit{gas} and \textit{gas-price}. 
Implementing a Smart Contract use case can pose a few security challenges because of the primary characteristic of BT of transparency. Moreover, the immutability of BT makes patching the already deployed Ethereum Smart Contracts to overcome the discovered vulnerabilities impossible.
\subsection{Solidity}
\vspace{-0.1 cm}
Ethereum Smart Contracts \cite{Ethereum} are typically written in a high-level turing-complete programming language such as Solidity\cite{Solidity}, and then compiled to the Ethereum Virtual Machine (EVM) byte-code\cite{Ethereum}, a low-level stack-based language. For instance, Listing \ref{HYIP} shows a smart contract written in the Solidity programming language\cite{Solidity}.

In Listing \ref{HYIP}, the first line of the program declares the Solidity version. The program will be compatible with the corresponding EVM or any other higher version less than 0.6.0. It contains a constructor to create an instance of the contract, and functions.

The \textit{msg.sender} is a built-in global variable representative of the address that is calling the function. The \textit{msg.value} is another built-in variable that tells us how much ether has been sent. The \textit{payable} keyword is unique to Solidity language as it allows a function to send and receive ether. 
 
A function with no name followed by \textit{payable} keyword, \textit{function () payable{}}, is a \textit{fallback} function in Solidity. This function gets triggered whenever a function call has an identifier that does not match any of the existing functions in a smart contract or if there was no data supplied to the function call at all. 

To transfer ether between the contracts, Solidity uses \textit{send()}, \textit{transfer()} and \textit{call()}. \textit{send()} transfers the ether and executes the fallback function of the contract that receives ether. The gas limit per execution is 2300 and an unsuccessful execution of \textit{send()} function does not throw an exception, but the gas required for the execution is spent. Similarly, \textit{transfer()} is also used to transfer ether between the contract, but the gas limit can be redefined using the \textit{.gas()} modifier, and an unsuccessful \textit{transfer()} throws an exception. The gas limit in both these functions prevents the security risk involved in executing expensive state changing code in the fallback function of the contract receiving the ether. The one pitfall is when a contract sets a custom amount of gas using the \textit{.gas()} modifier.
The \textit{call()} function is comparatively more vulnerable as there is no gas limit associated with this function.

\section{DoS caused by an Unexpected Revert Vulnerability}
When the flow of control is transferred to an external contract, the execution of the caller contract can fail accidentally or deliberately, which can cause a \textit{DoS} state in the caller contract. The caller contract can be in a \textit{DoS} state when a transaction is reverted due to a failure in the receiving smart contract, or the receiving smart contract deliberately reverts the transaction to disrupt the execution of the caller contract.
\subsection{DoS - Unexpected Revert Scenarios}

Scenario 1:
Consider an Ethereum Smart Contract \textit{HYIP}, which is a Ponzi scheme. This contract sends payments to lenders from funds collected via new lenders each day. The function \textit{sendPayment()} in Listing[\ref{HYIP}] contains the \textit{DoS due to an Unexpected Revert} vulnerability. The attack proceeds as follows: 
\begin{enumerate}
    \item The \textit{AttackerContract} lends funds to the \textit{HYIP} contract and throws an exception in its \textit{fallback} function.
\item When the function \textit{sendPayment()} is called to send daily payments to the contract lenders, the \textit{fallback} function of all its lenders are invoked and and the \textit{fallback} function of our \textit{AttackerContract} throws an exception, causing a deliberate revert of the transaction and subsequently transitioning the vulnerable contract \textit{HYIP} into a \textit{DoS} state.
\end{enumerate}

\begin{lstlisting}[language=Solidity, label={HYIP}, caption={Contract HYIP - Exploited for DoS-Unexpected Revert Vulnerability \cite{Etherscsan}}] 
pragma solidity ^0.5.0;

contract HYIP {
    Lenders[] private lender;
    function sendPayment() {
        for(uint i=lender.length; i<=0; i--) {
        uint payment=(lender[i].amount*/1000;
        if(!lender[i].addr.send(payment)) throw;
            }
    }
 }
 contract AttackerContract {
    bool private attack = true;
    function() payable {
        if (attack) throw; // callee fails the caller execution deliberately
    }
    function stopAttack() {
        if(msg.sender == owner) attack = false;
    }
 }
\end{lstlisting}

Scenario 2: 
Another scenario is when a contract uses iteration statements to \textit{push} out payments to users. The vulnerability arises when there is even one failed \textit{send}. For even one failed \textit{send} the whole payout system gets reverted, leaving the loop of payout system inoperable. No one gets the payment because one contract is causing a revert either accidentally or deliberately.(See Listing[\ref{loopPattern}])
\begin{lstlisting}[language= Solidity, caption={Iteration statements DoS Vulnerability Pattern},label={loopPattern}]
    address[] private paymentAddresses;
    mapping (address => uint) public payments;
    function payAll() public {
        for(uint x; x < paymentAddresses.length; x++) {
            require(paymentAddresses[x].send(payments[paymentAddresses[x]]));
        }
\end{lstlisting}
This scenario is more severe than the previous one as now a single payment failure in a \textit{send} will hold up all the other payments.
\subsection{Preventive Technique}
This vulnerability exists because of inadequate exception handling around conditional and iteration statements. Placing any external calls initiated by a callee contract into a separate transaction can be helpful in reducing the damage caused by this vulnerability. 

Making the recipient to \textit{pull} funds out rather than sender using \textit{push} to send out funds may help in avoiding this vulnerability. Moreover, isolating following code constructs should be considered to avoid this vulnerability:
\begin{enumerate}
    \item an if-statement with an external function call in the condition and a throw or a revert in the body;
    \item a for-statement with an external function call in the condition.
\end{enumerate}

\section{Motivation}

Although classified as \textit{not severe} by few researchers, a recent survey paper by Samreen and Alalfi \cite{Survey2020} highlights the importance of detecting a DOS vulnerability in an Ethereum Smart Contract. DOS vulnerability can render a smart contract inoperable thereby freezing any ether stored in the smart contract and also costs the Ethereum blockchain network in terms of memory and space which cannot be retrieved back in any way.
One of the real world instances of exploitation of this vulnerability is the \textit{GovernMental} smart contract. This was a Ponzi scheme which collected a fairly large amount of ether is an example of \textit{DOS due to unexpected revert} vulnerability in Ethereum Smart Contracts. It had collected almost 1100 ether at one point and was then susceptible to the \textit{DOS} vulnerability. This contract required sending a minimum amount to the previous creditors and then a deletion of these creditors in the scheme in order to withdraw the jackpot ether amount. However, as a part of clearing out the previously engaged creditors in this scheme, the contract iterates over the storage locations and deletes them one by one after sending out the minimum credit. And, in doing so, the \textit{fallback} function of all the creditors gets executed when there is a transfer of the minimum credit to their account. One of the creditor’s \textit{fallback} function reverted the transfer, thereby halting the entire scheme. Therefore, \textit{Governmental} has been stuck since then and its 1100 ETH are in a limbo.
The working of a Denial of Service vulnerability is tightly coupled with the error handling mechanism in Solidity-based Ethereum Smart Contracts. To identify this vulnerability, the patterns for error handling in Solidity needs to be analysed.
The common pattern for enforcing permissions in Solidity-based Ethereum Smart Contracts (till V 0.5.0) was to check the address of the account that is trying to access a smart contract  in an \textit{if} statement and to \textit{throw} if false.  
\vspace{3cm}
\begin{lstlisting}[language= Solidity, caption={If...Throw pattern of  DoS Vulnerability},label={ifThrowPattern}]
contract HasAnOwner {

    address owner;
    function restrictedFunction(){ 
        if (msg.sender != owner) { throw; }
        // do something only the owner should be allowed to do
    }
}
\end{lstlisting}
In the above listing \ref{ifThrowPattern}, if the \textit{restrictedFunction()} function is called by any other address other than owner of the contract, then the function will \textit{throw} and returns an invalid opcode error, undoing all state changes, and using up all remaining gas. 

The newer versions of Solidity use built-in functions like \textit{assert(), require()}, and \textit{revert()} to provide the same functionality, with a much cleaner syntax.
\begin{lstlisting}[language= Solidity,label={ifThrowPattern1}]
if(msg.sender != owner) { throw; }
\end{lstlisting}
currently behaves exactly the same as all of the following:
\begin{lstlisting}[language= Solidity, label={ifThrowPatternSimilars}]
if(msg.sender != owner) { revert(); }
assert(msg.sender == owner);
require(msg.sender == owner);
\end{lstlisting}
See Table \ref{Patterns} for a list of patterns identified to detect a \textit{DoS-Unexpected Revert} vulnerability. 

\begin{table}
\caption{DoS due to Unexpected Revert Vulnerability Patterns Identified}
\begin{center}
 \begin{tabular}{|p{0.5cm} p{2.5cm} p{4.25cm}|} \hline
\textbf{S.no.} &\textbf{Pattern Name} & \textbf{Pattern Format}\\
\hline
1. & Iteration Pattern  & \begin{lstlisting}[numbers=none]
for(uint x; x < rec.length; x++) {
        require(rec[x].send(payments[rec[x]]));} 
    \end{lstlisting}\\ \hline
2. & If/Throw Pattern & \begin{lstlisting}[numbers=none]
if(msg.sender != owner) { throw; }
\end{lstlisting}\\ \hline
3. & If/Revert Pattern & \begin{lstlisting}[numbers=none]
if(msg.sender != owner) { revert(); }
\end{lstlisting} \\ \hline
4. & Assert Pattern & \begin{lstlisting}[numbers=none]
assert(msg.sender == owner);
\end{lstlisting}\\ \hline
4. & Require Pattern & \begin{lstlisting}[numbers=none]
require(msg.sender == owner);
\end{lstlisting}\\ \hline
\end{tabular}
\end{center}
\label{Patterns}
\end{table}

\begin{figure}[!ht]
\centerline{\includegraphics[width=.42\textwidth]{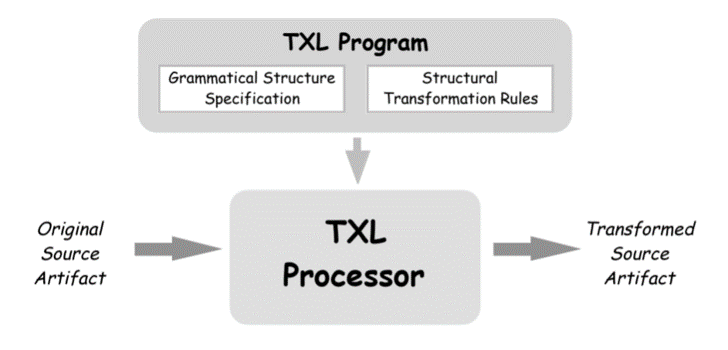}}
\caption{Txl Paradigm\cite{TXL}}
\vspace{-0.5 cm}
\label{fig:txl}
\end{figure}
\section{Proposed Framework}
There are two parts to check for the DOS due to unexpected revert vulnerability. First, a call to an external function is executed in a module of the smart contract that could possibly mishandle an exception if thrown. Second, there is an exception thrown at the external call causing the denial of service state of the smart contract. 
Therefore, the pattern in our proposed analysis solution, SmartScan,  should combine both static and dynamic analysis methodologies to check the possibility of an exception being thrown at an external call and to confirm the functional unavailability of the smart contract after the external call to an untrustworthy address dynamically, i.e. mishandled exception causing DOS due to an unexpected revert by the untrustworthy address.
\begin{figure*}[h]
\centerline{\includegraphics[width=1\textwidth]{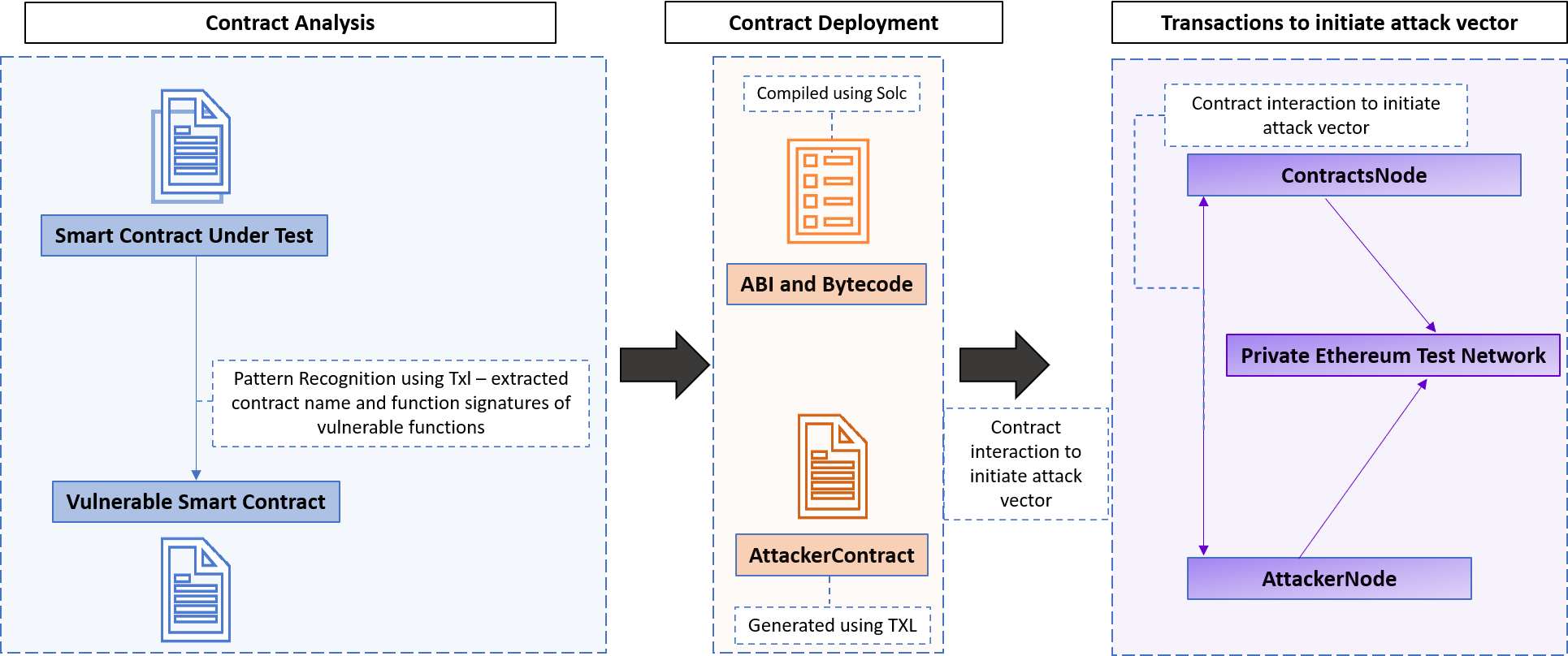}}
\caption{Proposed Framework}
\label{fig:architecture}
\end{figure*}
This framework utilizes both static and dynamic analysis techniques to cover almost all the above-described attack patterns. The first stage of SmartScan adopts static analysis technique of pattern recognition and automated instrumentation of a Smart Contract Under Test (SCUT) to efficiently cover the possible attack scenarios in the SCUT.

The second stage uses the extracted information of vulnerable functions in the SCUT from stage one to automatically generate an attacker-contract which is then used to interact with the SCUT after they are deployed onto our private Ethereum test network. This interaction is done to demonstrate the exploitability  of the \textit{DoS-Unexpected Revert} vulnerability in a SCUT dynamically. 

\subsection{Stage 1: Contract Analysis}
To accurately model the possible scenarios that can result in this vulnerability, our proposed framework, SmartScan [as shown in Figure \ref{fig:architecture}], uses Txl\cite{TXL}. TXL is a programming language used for structural analysis and source transformation. The TXL paradigm 
consists of a grammar and transformation rules that are used to parse a given input text into Abstract Syntax Tree (AST), and this intermediate AST is transformed into a new AST of the target domain, and finally this output is un-parsed to a new output text.

The structure of a transformation rule in TXL specifies a target program type to be transformed, a pattern which is an instance of the type that we are interested in replacing and a replacement structure. In SmartScan, we used TXL to achieve the following tasks: 
\begin{enumerate}
\item To parse the SCUT and to apply pattern recognition rules in order to identify potentially vulnerable functions in the SCUT,
\item To extract the names of these potentially vulnerable functions from the SCUT to generate an attacker-contract. 
\end{enumerate}
\begin{lstlisting}[language= Solidity, caption={Txl program structure},label={lst6}]
 rule main
 replace [program]
 EntireInput [program]
    by
     EntireInput [Function1]
                 [Function2]
 end rule
\end{lstlisting}
The potentially vulnerable functions were determined if they contained an external call using any of these three Solidity’s built-in functions: \textit{transfer()}, \textit{call()}, \textit{send()} in an iterative or a conditional statement or a statement that could mishandle an exception if thrown. These exception mishandling prone statements include: \textit{assert(), revert()}.  These function signatures were extracted using TXL and outputted in a .txt file. This output file was used to modify the \textit{attacker-contract} to exploit \textit{DOS} vulnerability in the SCUT by deliberately reverting the transaction in the \textit{fallback} function of the attacker-contract.


An \textit{attacker-contract}, as shown in Listing\ref{lst8}, is automatically created to interact with the potentially vulnerable functions of the SCUT to recreate the DOS due to unexpected revert attack scenario. For illustration, \textit{BetDeEx} smart contract is used as an example from the collected data-set for this research. This framework successfully detected \textit{DOS} vulnerability within it in our analysis.

As shown in Listing\ref{lst8}, the \textit{AttackerContract} deliberately reverts the transaction using \textit{revert()} in its \textit{fallback} function. At first, the attacker-contract creates an instance of the \textit{BetDeEx} contract and thus calls the constructor of the \textit{BetDeEx} contract. It then calls the \textit{collectPlatformFee()} function of \textit{BetDeEx} smart contract to initiate the attack. Within the \textit{collectPlatformFee()} function, the \textit{BetDeEx} smart contract transfers ether using the in-built Solidity function \textit{transfer()} in a \textit{require()} statement. Since the transfer function call has no parameters provided, it will invoke the \textit{fallback} function of the \textit{attacker-contract}. If the transaction fails, then an exception is thrown by the \textit{require()} statement. Due to the mishandling of this exception in the \textit{require()} statement, the \textit{BetDeEx} is rendered unavailable for future requests. 
\begin{lstlisting}[language= Solidity, caption={Attacker-Contract for BetDeEx SCUT},label={lst8}]
import "BetDeEx.sol";
contract Attacker_BetDeEx{
    address payable private _owner;
    address payable private _vulnerableAddr;
    BetDeEx public fd = BetDeEx(_vulnerableAddr);
    constructor() public {
         _owner = msg.sender;
         fd. collectPlatformFee();
    }
    function() external{
        revert();
    }}
\end{lstlisting}
Within the \textit{fallback} function, the \textit{attacker-contract} can deliberately have a \textit{revert();} statement to cause a revert of any transaction happening between this attacker-contract and any other contract. As a result, the \textit{BetDeEx} smart contract will be in a state of unavailability due to mishandling of this unexpected revert caused in an external call. 
With the help of the \textit{attacker-contract}, we can verify the exploitability of the potential \textit{DOS due to unexpected revert} vulnerability detected by static analysis using TXL pattern matching.
\subsection{Stage 2: Contracts Deployment}
\vspace{-0.15 cm}
The contracts deployment stage requires an Ethereum Private Test Network with test environment specifications. To deploy the SCUT and the \textit{attacker-contracts} for testing, we used Geth\cite{Geth}, a Golang implementation of Ethereum, to create a private blockchain. This enabled us to create a new and private blockchain which we will be using to deploy and test any smart contract. This private blockchain does not connect to the Ethereum Main-net and therefore does not require real ether to execute transaction between smart contracts.
\subsubsection{Geth}
To use Geth\cite{Geth}, we created a new account that represents a key pair. We simulated having two different accounts that store our private blockchain by creating two nodes in the blockchain. These nodes serve the purpose of acting like two different computers hosting and interacting with the same private blockchain using two terminal windows. We used two data directories, so each node has a separate place to store their local copy of our blockchain. 
\subsection{Stage 3: Initiating Attack Vector}
To interact with the blockchain, we launched the geth \textit{console} command to initiate the ports for ContractsNode and AttackerNode. 

After simulating the contracts node and attacker node on our private Ethereum network, the SCUTs are deployed on these nodes in the following steps:
\begin{enumerate}
    \item The SCUT were compiled and deployed using a Javascript in which the \textit{Solc} command was invoked to generate the ABI and bytecode of the SCUTs.(See Listings \ref{JavaScriptCompile} and \ref{JavaScriptDeploy}) 
    
    \item \textit{HDWalletProvider} is a hosted Ethereum node cluster that was used to automatically deploy the contracts onto our Private Ethereum Test Network.
    
    \item After deploying the contract onto a node, the transaction is submitted to all the peers of the blockchain. The \textit{miner.start()} command starts the mining process and \textit{miner.stop()} ends this process. After the mining process is completed, the contract gets appended to the blockchain. 
\end{enumerate}
\vspace{-0.3cm}
\begin{lstlisting}[language= Solidity, caption={Javascript to Compile all the SCUTs automatically},label={JavaScriptCompile}]
    const compileContracts = () => {
    try{
		const compiledContracts = JSON.parse(solc.compile(JSON.stringify(input)).contracts);
		if (!compiledContracts) {
        console.error('>> ERRORS <<<\n', 'NO OUTPUT');
    } else if (compiledContracts.errors) { // something went wrong.
        console.error('>> ERRORS <<\n');
        compiledContracts.errors.map(error => console.log(error.formattedMessage));
    }
	for (let contract in compiledContracts) {
		for(let contractName in compiledContracts[contract]) {
			fs.outputJsonSync(
				path.resolve(builPath, `${contractName}.json`),
				compiledContracts[contract][contractName],
				{spaces: 2})
				}}} 
	catch(e){
		console.log(e);}}
\end{lstlisting}
\vspace{0.9 cm}
 \begin{lstlisting}[language=Solidity, caption={Javascript to Deploy all the SCUTs automatically},label={JavaScriptDeploy}]
    (async () => {
	const accounts = await web3.eth.getAccounts();

	console.log(`Attempting to deploy from account: ${accounts[0]}`);

	const deployedContract = await new web3.eth.Contract(compiledContract.abi)
		.deploy({
			data: '0x' + compiledContract.evm.bytecode.object,
			arguments: [3, 5]
		})
		.send({
			from: accounts[0],
			gas: '2000000'
		});

	console.log(
		`Contract deployed at address: ${deployedContract.options.address}`
	);

	provider.engine.stop();
})();
\end{lstlisting}

After deploying the contracts, transactions were initiated between SCUTs and the \textit{attacker-contract} using the vulnerable function name extracted using TXL in the first step of our analysis. 








\begin{figure*}[!t]
\centerline{\includegraphics[width=1\textwidth]{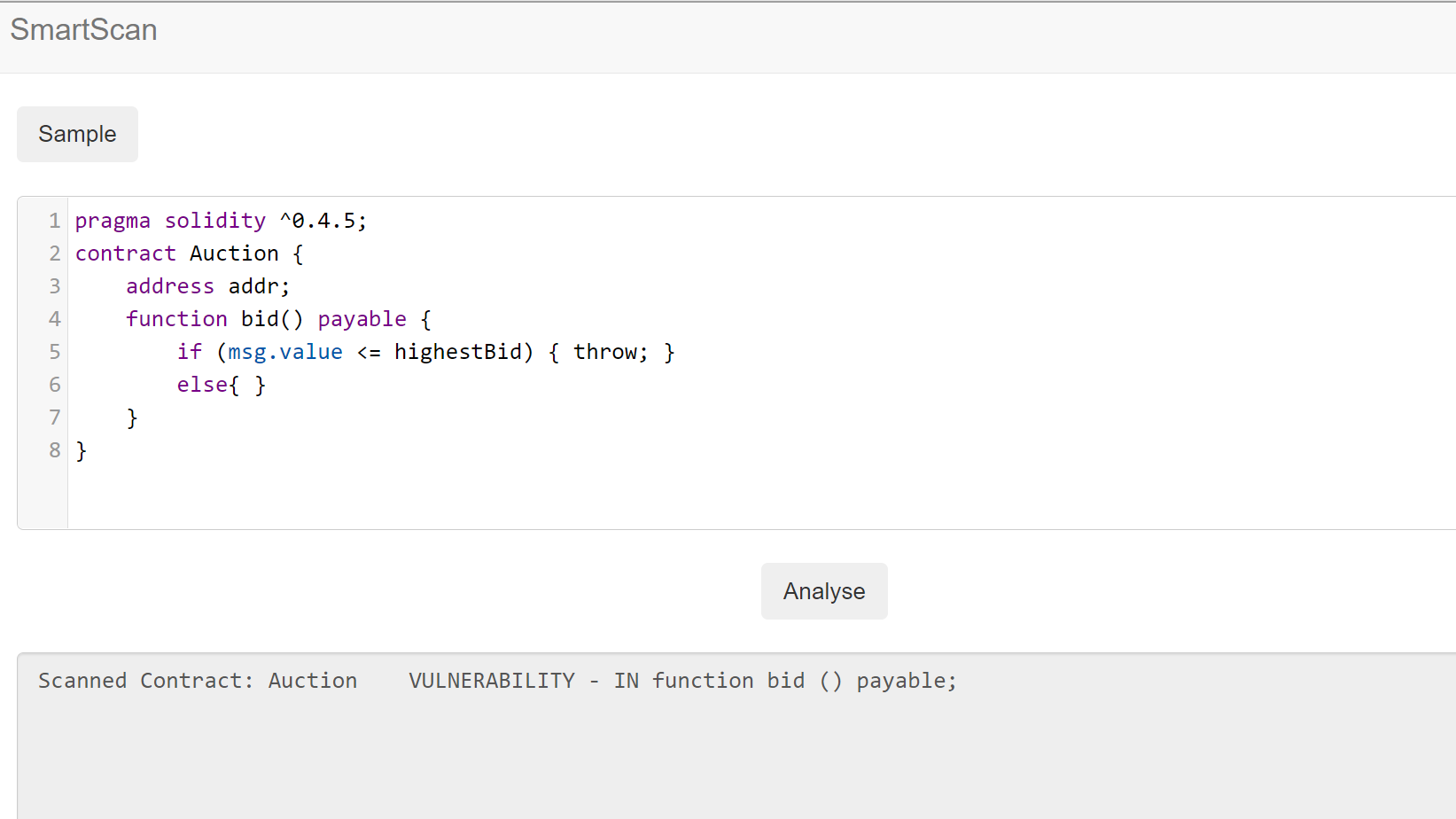}}
\caption{Report of Analysis Generated}

\label{fig:smartscan}
\end{figure*}

\section{Usage of Our Tool}
We deployed SmartScan as a web service. 
To analyze a smart contract, our tool adopts a one-click design.
(See Figure[\ref{fig:smartscan}]). 
Users only need to type the contract code in the browser, and our tool will then statically analyse and give the results as \textit{Passed} if there are no potentially vulnerable functions detected in the SCUT.

However, if there are potentially vulnerable functions in the SCUT, then the tool deploys the SCUT onto our private Ethereum Test Network and begins the second stage of our analysis i.e. it generates an attack vector to confirm the exploitability  of the DOS due to unexpected revert vulnerability in the SCUT. 

Taking the simple SCUT in Figure [\ref{fig:smartscan}] as an example, we briefly describe the usage of our tool. By clicking the \textit{Scan} button, our tool initiates the analysis. When the analysis is done, a report is generated as shown in Figure[\ref{fig:smartscan}]. In this case, the potentially vulnerable function signatures are highlighted.

Access our tool here, SmartScan \footnote{https://smartscan.cs.ryerson.ca:4638/}.

\section{Related Work}

Many tools have been developed that aim at improving the security and correctness of Ethereum smart contracts. The online Solidity compiler Remix also provide warnings and suggestions regarding some of the known vulnerabilities in Smart Contracts. However, there is very few tools that has been developed to check for DOS due to unexpected revert vulnerability.

Grech et. al developed a tool, MadMax (now renamed as ContractLibrary) \cite{madmax} as a static program analysis tool that aims at automatically detecting gas-focused vulnerabilities. This tool combines a control-flow-analysis-based decompiler, Vandal\cite{Vandal} and declarative program-structure queries.  Grech et. al proposes that this combined analysis achieves high precision and scalability. ContractLibrary \cite{madmax} outlines the DOS-due to unexpected revert as \textit{Wallet Griefing} - a condition in which a contract may throw an out-of-gas exception when combined with other complicating factors.

SMARTCHECK is a pattern-based analysis tool that uses XPath to detect if any vulnerabilities pattern exists in a Smart Contract. To do so, it transforms the Smart Contract into XML representation. SmartCheck has its limitations, as detection of this vulnerability requires a combination of static and dynamic analysis techniques to confirm the unavailable state of a vulnerable Smart Contract. 

SMARTSHIELD by Y. Zhang et al. \cite{smartshield} utilizes EVM bytecode analysis and provides an automatic correction mechanism that would avoid vulnerable patterns in Ethereum smart contracts. It does this rectification by extracting EVM bytecode level semantic data to transform the vulnerable smart contracts into secure ones.

ETHPLOIT by Q. Zhang et al. \cite{Ethploit} is another tool that addresses this vulnerability. ETHPLOIT automatically detect vulnerabilities that have been exploited in Ethereum smart contracts. This tool adopts light-weight techniques to answer the problems of previous such tools. These problems consisted of unsolvable constraints and Blockchain effects. It is claimed by Q. Zhang et al. \cite{Ethploit} that this tool achieves precise and efficient smart contract analysis and successfully detects more exploits than previous exploit generation tools.
A recent study published in Feb. 2020 by Durieux et al. aims at reviewing the findings of automated tools available for detecting known vulnerabilities in Ethereum Smart Contracts. The empirical study by Durieux et al. lists two other automated tools Osiris\cite{osiris} and Slither\cite{slither} that detect DOS vulnerabilities. 

However, we found that Osiris \cite{osiris} focuses on identifying integer overflow vulnerabilities in Ethereum smart contracts and checks for semantics of a smart contract that may cause an integer overflow vulnerability by applying symbolic execution and taint analysis. 

Slither\cite{slither} provides a way to identify if the \textit{send} built-in function of Solidity was used without checking its return value. This detection is a much broader way of formulating pattern for the occurrence of  DOS vulnerability and Slither fails to narrow down the pattern of function calls that may cause this vulnerability. 
\section{Evaluation}
The data-set used for the evaluation consisted of 500 real world Smart Contracts of Solidity version greater than 0.5. This was extracted from Etherscan by developing a python-based web scraper. Etherscan\cite{Etherscsan} is a free-to-use platform for BT analytics based on Ethereum. It is essentially a Block Explorer that allows users to easily lookup, confirm and validate transactions that have taken place on the Ethereum Blockchain. Ethereum Smart Contracts developers can get a substantial benefit from APIs available that can be utilized to either build decentralized applications or serve as a data-set for Ethereum Blockchain analysis.

Although we mentioned that four automated testing tools are available to detect the \textit{DOS- due to unexpected revert} vulnerability (SMARTSHIELD \cite{smartshield}, ETHPLOIT \cite{Ethploit}, SmartCheck \cite{SmartCheck} and ContractLibrary \cite{madmax}). We could evaluate our framework - SmartScan against the results produced by only two of the available tools - SmartCheck and ContractLibrary, as these were the only publicly available tools to identify DOS vulnerabilities in Ethereum Smart Contracts.

The analysis result shows that 73 contracts from the 500 contracts were deployed onto our private Ethereum Test Network to confirm the existence of DOS due to unexpected revert vulnerability as they exhibited the identified vulnerable code patterns. In the second stage of our analysis, out of the deployed 73 contacts, only 28 were confirmed to be exploitable by a DOS attack. Further breakdown of these 28 confirmed vulnerabilities indicated that 18 of these confirmed vulnerable SCUTs were exploitable by initiating an attack vector from an external contract. The remaining 10 vulnerable SCUTs were protected by an access modifier and was successfully exploited by gaining control of the owner contract or its dependent contracts. 

Our framework, SmartScan, resulted in detection of \textit{DoS} vulnerability in 10 more smart contracts than SmartCheck\cite{SmartCheck} and ContractLibrary\cite{madmax} [See Table \ref{resultAnalysis}].
These two tools used in the evaluation failed to detect these externally non-exploitable vulnerabilities even though they contained the pattern that resulted in a \textit{DoS} state in a smart contract. This was due to the fact that the functions that exhibited the pattern for \textit{DoS-Unexpected Revert} vulnerability was protected by an access modifier that restricted its external invocation. We confirmed the existence of the DOS vulnerability in these smart contracts by invoking the vulnerable functions from within the SCUT itself.  
\begin{lstlisting}[language= Solidity, caption={Contract Unipool - for DoS-Unexpected Revert Vulnerability\cite{Etherscsan}},label={Unipool}] 
contract Unipool {
    function sendValue(address payable recipient, uint256 amount) internal {
        require(address(this).balance >= amount, "Address: insufficient balance");
        if(!recipient.call.value(amount)(""))
        throw;
    }
}
\end{lstlisting}

For example, one of the smart contracts in our dataset - \textit{Unipool}\cite{Etherscsan},listing \ref{Unipool} 
had an access modifier \textit{internal} for a function \textit{sendValue} which made it inaccessible to be called from an external contract. But the pattern that exposes the smart contract to a possible DOS-due to unexpected revert from an external contract did exist in the function \textit{sendValue} of the SCUT \textit{Unipool}. Therefore, the confirmation of the DOS vulnerability in this case could be determined by trying to gain control of the owner contract or the dependent contracts of the SCUT. It can serve as an indication to the smart contract programmers that such coding practice in developing smart contracts may expose their smart contracts to a the DOS-due to unexpected revert vulnerability. 

Our framework, SmartScan, produced no false positives and therefore, has better precision than SmartCheck \cite{SmartCheck}. ContractLibrary \cite{madmax},however, had similar precision value when compared to our framework, SmartScan, but lower recall value because of missed vulnerability instances detection (false negatives). Also, the process of checking a smart contract for vulnerabilities using ContractLibrary is complicated as it requires deploying the smart contract under test (SCUT) on the Ropsten test network of the Ethereum Blockchain and then wait for ContractLibrary framework to import and test the SCUT. The process of detecting a smart contract for vulnerabilities is quicker and simpler in our framework, SmartScan.

\begin{table}
\caption{Results of \textit{DOS due to Unexpected Revert} vulnerability analysis. FN: False Negative}
  \begin{center}
 \begin{tabular}{|p{2.2cm} p{1.4cm} p{1.65cm} p{2.2cm}|} 
 \hline
 \textbf{Analysis Results} &  \textbf{SmartScan} &  \textbf{SmartCheck} & \textbf{Contract Library} \\ 
 \hline\hline
   Vulnerabilities Confirmed & 28 & 20 & 18  \\
 \hline
  Exploitable-internal attack & 10/28 & 0/20 & 0/18     \\
  \hline 
  False-Positives (FP) & 0/28 & 2/20 & 0/18 \\
  \hline
  Vulnerability pattern instances missed (FN) & 0 & 10 & 10\\
  \hline
  \hline 
\end{tabular}
\end{center}
\label{resultAnalysis}
\end{table}

\begin{table*}
\caption{Detailed Evaluation of \textit{DOS due to Unexpected Revert} vulnerability analysis. FP: False Positive; FN: False Negative; {\color{blue}{\cmark}}: Vulnerability found; {\color{red}{\xmark}}: Vulnerability not found; {\color{brown}{\warning}}: In-exploitable due to inaccessibility }
\begin{center}
 \begin{tabular}{|p{0.7cm} | p{4cm} | p{2.5cm} | p{5cm} | p{2cm} | p{2.5cm}|}
 \hline
\textbf{S.no.} & \textbf{Smart Contract} & \multicolumn{2}{|c|}{\textbf{SmartScan}} & \textbf{SmartCheck} & \textbf{ContractLibrary}\\
\hline
\hline
&&\textbf{Pattern Detected} & \textbf{Exploitable by external Attacker Contract}&&\\
\hline
\hline
1. & AZBICore & {\color{blue}{\cmark}} If/Throw  & {\color{blue}{\cmark}} & \color{blue}{\cmark}&\color{blue}{\cmark}\\\hline
2. & AZR & {\color{blue}{\cmark}} Require & \color{blue}{\cmark} & \color{blue}{\cmark}&\color{blue}{\cmark}\\\hline
3. & BetDeEx & {\color{blue}{\cmark}} Iteration & \color{blue}{\cmark} & \color{blue}{\cmark}&\color{blue}{\cmark}\\\hline
4. & BIOXE &  {\color{blue}{\cmark}} If/Throw & \color{blue}{\cmark} & \color{blue}{\cmark}&\color{blue}{\cmark}\\\hline
5. & Booth Renting &  {\color{blue}{\cmark}} Assert & \color{blue}{\cmark} &\color{blue}{\cmark}&\color{blue}{\cmark} \\\hline
6. & Chainlink &  {\color{blue}{\cmark}} If/Throw & \color{blue}{\cmark} & \color{blue}{\cmark} &\color{blue}{\cmark}\\\hline
7. & CREDIT &  {\color{blue}{\cmark}} If/Throw & \color{blue}{\cmark} & \color{blue}{\cmark}&\color{blue}{\cmark}\\\hline
8. & CryBet &  {\color{blue}{\cmark}} Iteration  & \color{blue}{\cmark} & \color{blue}{\cmark}&\color{blue}{\cmark}\\\hline
9. & ETGF &  {\color{blue}{\cmark}} Require  & \color{blue}{\cmark} & \color{blue}{\cmark}&\color{blue}{\cmark}\\\hline
10. & Flash Loan Ave &  {\color{blue}{\cmark}} If/Throw & \color{blue}{\cmark} & \color{blue}{\cmark}&\color{blue}{\cmark}\\\hline
11. & Lition Pool &  {\color{blue}{\cmark}} Iteration & \color{blue}{\cmark} & \color{blue}{\cmark}&\color{blue}{\cmark}\\\hline
12. & Load Coin &  {\color{blue}{\cmark}} Iteration & \color{blue}{\cmark} & \color{blue}{\cmark}&\color{blue}{\cmark}\\\hline
13. & RICH &  {\color{blue}{\cmark}} Assert & \color{blue}{\cmark} & \color{blue}{\cmark}&\color{blue}{\cmark}\\\hline
14. & Syndicate &  {\color{blue}{\cmark}} Iteration & \color{blue}{\cmark} & \color{blue}{\cmark} &\color{blue}{\cmark} \\\hline
15. & Tip &  {\color{blue}{\cmark}} Iteration & \color{blue}{\cmark} & \color{blue}{\cmark} & \color{blue}{\cmark}\\ \hline
16. & USPi &  {\color{blue}{\cmark}} Require  & \color{blue}{\cmark} & \color{blue}{\cmark} & \color{blue}{\cmark}\\\hline
17. & HYIP &  {\color{blue}{\cmark}} If/Throw & \color{blue}{\cmark} & \color{blue}{\cmark}& \color{blue}{\cmark}\\\hline
18. & CallToTheUnknown &  {\color{blue}{\cmark}} Require & \color{blue}{\cmark} & \color{blue}{\cmark}& \color{blue}{\cmark} \\\hline
&&&\textbf{Inaccessible from external contract because pattern is protected with an access modifier}& & \\\hline
19. &  BKRW &  {\color{blue}{\cmark}} Assert & {\color{brown}{\warning}}Access modifier-\textit{only owner} & {\color{red}{\xmark}} (FN)& {\color{red}{\xmark}} (FN)\\\hline
20. & GramChain &   {\color{blue}{\cmark}} If/Revert & {\color{brown}{\warning}}Access modifier-\textit{only owner} & {\color{red}{\xmark}} (FN)&{\color{red}{\xmark}} (FN) \\\hline
21. & Price Feed &  {\color{blue}{\cmark}} Assert & {\color{brown}{\warning}} Access modifier-\textit{internal} & {\color{red}{\xmark}} (FN)& {\color{red}{\xmark}} (FN)\\\hline
22. &  Off-Chain Asset Evaluator &   {\color{blue}{\cmark}} If/Throw & {\color{brown}{\warning}} Access modifier-\textit{internal} & {\color{red}{\xmark}} (FN) & {\color{red}{\xmark}} (FN) \\\hline
23. &  Off-Chain Currency Evaluator &  {\color{blue}{\cmark}} If/Throw & {\color{brown}{\warning}}Access modifier-\textit{internal} & {\color{red}{\xmark}} (FN)&{\color{red}{\xmark}} (FN) \\\hline
24. & Redeem Code &  {\color{blue}{\cmark}} Require & {\color{brown}{\warning}}Access modifier-\textit{only owner} & {\color{red}{\xmark}} (FN)& {\color{red}{\xmark}} (FN)\\\hline
25. & SONERGY &  {\color{blue}{\cmark}} Assert  & {\color{brown}{\warning}}Access modifier-\textit{only owner}& {\color{red}{\xmark}} (FN)& {\color{red}{\xmark}} (FN)\\\hline
26. & Unipool &  {\color{blue}{\cmark}} Require & {\color{brown}{\warning}}Access modifier-\textit{internal} & {\color{red}{\xmark}} (FN)& {\color{red}{\xmark}} (FN)\\\hline
27. & Xank &  {\color{blue}{\cmark}}  Require & {\color{brown}{\warning}}Access modifier-\textit{internal} & {\color{red}{\xmark}} (FN)& {\color{red}{\xmark}} (FN) \\\hline
28. & ZdzMedia &  {\color{blue}{\cmark}} Assert & {\color{brown}{\warning}}Access modifier-\textit{only owner} & {\color{red}{\xmark}} (FN) &{\color{red}{\xmark}} (FN) \\\hline
29. &  Countdown Griefing &  \color{red}{\xmark} &  & {\color{blue}{\cmark}} (FP) & {\color{red}{\xmark}}\\\hline
30. &  Simple Griefing&  \color{red}{\xmark} &  & {\color{blue}{\cmark}} (FP) & {\color{red}{\xmark}}\\\hline
\hline
\end{tabular}
\end{center}
\label{ContractsDet}
\end{table*}

\section{Conclusion}
In this paper, a combined static and dynamic analysis framework is proposed to detect \textit{DoS-Unexpected Revert} vulnerabilities in data-set of 500 real world Ethereum Smart Contracts collected from Etherscan\cite{Etherscsan}. These Smart Contracts were collected by developing a python-based web scraper for Etherscan portal. This framework uses TXL\cite{TXL} for analyzing these smart contracts to extract the function signature of potentially vulnerable functions in a contract.
Then, the \textit{DoS} vulnerability confirmation is done at the run-time while interacting with the vulnerable contract using an \textit{attacker-contract} in an attempt to recreate the \textit{DoS} scenario.
\section{Future Work}
The future work in this research would be to make our tool scalable to include Solidity smart contracts of newer versions. Currently, this framework targets Solidity smart contracts of versions v5.0.
Also, to include detection of other vulnerabilities listed by some of the known security vulnerabilities research\cite{Survey2020} by our framework, SmartScan. 
\section*{Acknowledgments}
This work is supported in part by the Natural Sciences and Engineering Research Council of Canada (NSERC).
\section{Data Availability}
The data collected to conduct this research is publicly available on Etherscan web portal and was gathered using a web scraper written in Python.
The tool produced in this research to analyse smart contracts for the presence of DOS due to unexpected revert vulnerability is also publicly available at SmartScan \footnote{https://smartscan.cs.ryerson.ca:4638/}.
\clearpage
\balance
\bibliographystyle{IEEEtran}
 \bibliography{Noama.bib}
\nocite*
\end{document}